\documentclass[aps,preprint,pre]{revtex4-1}
\usepackage[utf8]{inputenc}
\usepackage{float}
\usepackage[fleqn]{amsmath}
\usepackage[title]{appendix}
\usepackage{graphicx}
\usepackage{amsmath,geometry,amssymb,verbatim,caption,makecell,subcaption,tikz,color,hyperref,float}
\usepackage{geometry}
\usepackage{enumitem}
\usepackage{siunitx}
\begin{document}

\title{Dipolar Ising Model: Phases, Growth Laws and Universality}

\author{Shikha Kumari}
\author{Sanjay Puri}
\affiliation{School of Physical Sciences, Jawaharlal Nehru University, New Delhi - 110067, India}
\author{Varsha Banerjee}
\affiliation{Department of Physics, Indian Institute of Technology Delhi, New Delhi - 110016, India}

\begin{abstract}
The behavior of many magnetic and dielectric solids, and the more contemporary magnetic super-lattices, is governed by dipolar interactions. They are anisotropic and long-ranged, having varied consequences ranging from ground states with complicated magnetic order to the presence of glassy dynamics characterized by a plethora of relaxation times. These systems are well-captured by the dipolar Ising model (DIM) with nearest-neighbor exchange interactions ($J$) and long-range dipolar interactions ($D$). Depending on the relative interaction strength $\Gamma = J/D$, there are {\it four} phases of distinct magnetic order and symmetry. Using Monte Carlo simulations, we perform deep quenches to study domain growth or {\it coarsening} in the $d=3$ DIM. This important non-equilibrium phenomenon has not been addressed as dipolar interactions are notoriously difficult to handle theoretically. Our study reveals that, in spite of the anisotropy in interactions and diversity in ground state configurations, we observe {\it universality} in the ordering dynamics of {\it all} phases.
\end{abstract}

\maketitle

\section{Introduction}

Dipolar interactions are prevalent in magnetic and dielectric solids composed of rare-earths and transition metals \cite{kretschmer, merz, molnar, chakraborty2004, chamberlain2005, biltmo09, wu2017,cai2019}. They are anisotropic and long-ranged, and arise from nuclear magnetic moments in alkali hydrides and solid $^3He$,  electron magnetic moments in rare earth fluorides, chlorides and hydroxides, electric dipole moments in ferroelectric and anti-ferroelectric structures, etc. A large class of these compounds is well-described by the nearest-neighbor (nn) Ising model with dipolar interactions or the {\it dipolar Ising model} (DIM). For $N$ Ising spins on a $d$-dimensional cubic lattice with sites labeled by $i$, the Hamiltonian is given by
\begin{equation}
\mathcal{H} = - J \sum_{\langle ij \rangle} \sigma_i \sigma_j - D \sum_{\substack{{i,j} \\ i \neq j}}\left( \frac{3\cos^2{\theta}_{ij} - 1}{r_{ij}^3}\right) \sigma_i \sigma_j , \quad \sigma_i = \pm 1 .
\label{eq:ham}
\end{equation}
The first term on the right-hand-side is the contribution from the nn exchange energy. The parameter $J$ represents the interaction strength, and favors ferromagnetic (antiferromagnetic) alignment of spins for $J>0$ ($J<0$). The second term is the contribution from the dipole-dipole interactions, whose strength is given by $D = \mu^2/a^3$. Here, $\mu$ is the dipole moment of the spin, and $a$ is the spacing between nn sites of the underlying lattice. Further, $\vec{r}_{ij}$ is a vector joining sites $i$ and $j$ in units of lattice spacing $a$, and $\theta_{ij}$ is the angle made by $\vec{r}_{ij}$ with the Ising axis ($z$-axis). The presence of $r_{ij}^{3}$ in the denominator makes dipolar interactions long-ranged due to which spin-spin interactions are significant up to multiple lattice spacings. Consequently, the sum in the second term extends over all spin pairs. In addition to being long-ranged, the dipolar interactions are anisotropic, fluctuating in sign and strongly influenced by the underlying lattice structure. The $\theta_{ij}$-dependence implies that dipolar interactions can be zero, positive or negative -- depending on the positions of the spins $i$ and $j$. For a reference spin $i$, the interaction with spin $j$ is zero for $3\cos^2{\theta}_{ij} - 1=0$, antiferromagnetic for $\SI{55}{\degree} <\theta_{ij} <\SI{125}{\degree}$, and ferromagnetic for other values. Therefore, a ferromagnetic alignment of spins is favored along the $z$-direction, but domain walls are preferred along the $xy$-plane. 

\begin{figure}
\begin{center}
\includegraphics[width=14cm]{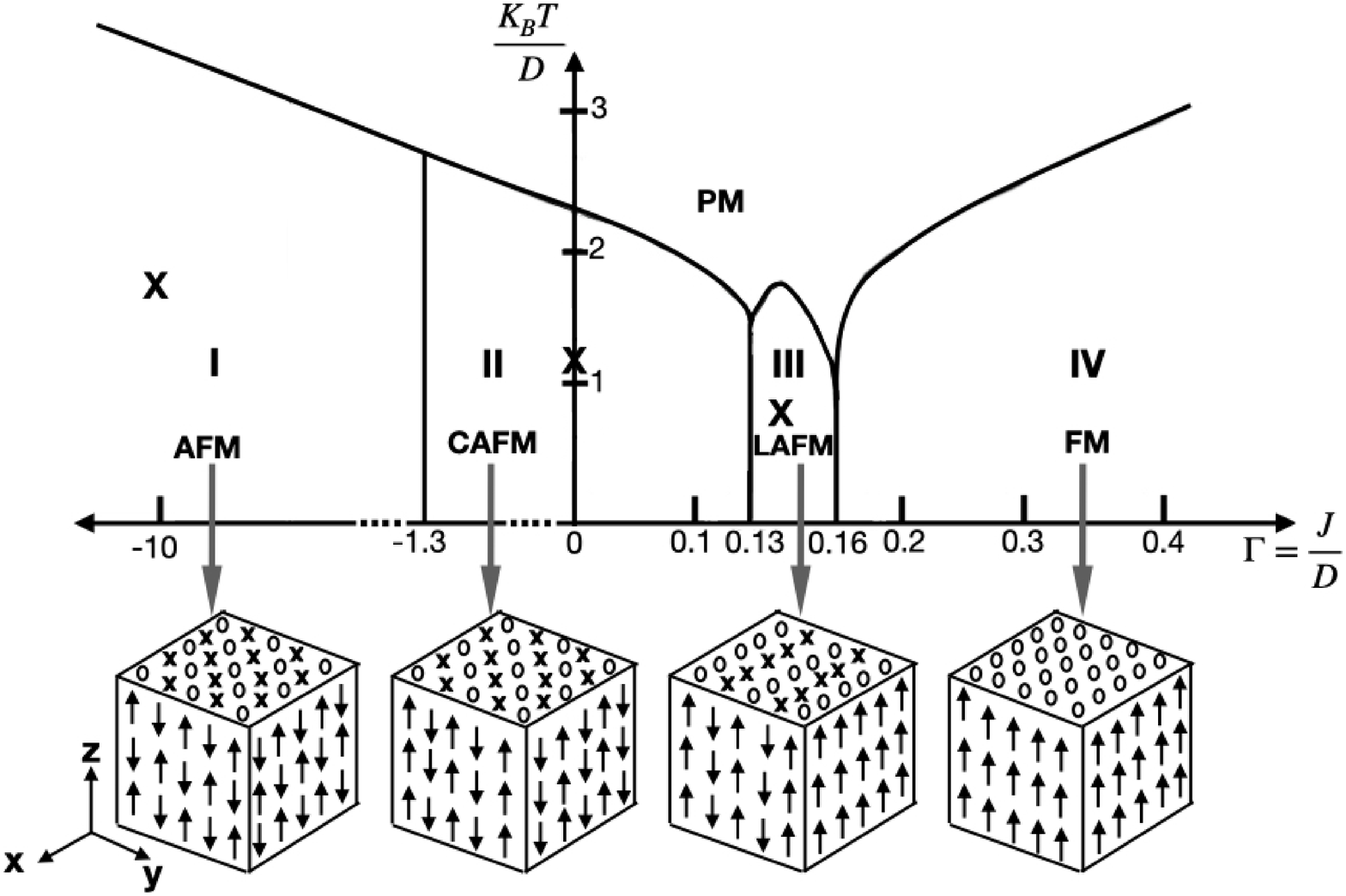}
\end{center}
\caption{Schematic phase diagram of the DIM model in the $(\Gamma, k_BT/D)$-plane, where $\Gamma = J/D$. We set $D=1$ and vary $J,T$. There are four distinct phases: (I) Antiferromagnet (AFM);  (II) Columnar antiferromagnet (CAFM);  (III) Layered antiferromagnet (LAFM); (IV) Ferromagnet (FM). The corresponding ground state configurations are shown below. The quench locations are marked by crosses, and satisfy $T=0.5T_c(\Gamma)$. The corresponding parameter values are $(\Gamma, k_BT/D)=(-10,1.75)$ for AFM; $(0,1.19)$ for CAFM; $(0.14,0.825)$ for LAFM; and $(\infty,\infty)$ for FM.}
\label{fig1}
\end{figure}

A phase diagram of the DIM was obtained by Kretschmer and Binder for a simple cubic lattice $(L^3, \ L=6,8)$ using Monte Carlo (MC) simulations \cite{kretschmer}. The interplay of the nn exchange interactions and the complicated dipolar interactions reveals rich phase behavior. The phase diagram in Fig.~\ref{fig1} has been obtained by varying $J, T$ with fixed $D=1$. Depending on the relative interaction strength $\Gamma = J/D$, the system exhibits four phases with distinct magnetic order and symmetry:\\
(I) For $\Gamma < -1.338$, the nn exchange interaction with $J<0$ dominates over the dipolar term. As expected, the ground state (GS) is an antiferromagnet (AFM).\\
(II) For $-1.338 < \Gamma < 0.127$, the dipolar term dominates over the nn exchange interaction. The ground state is anisotropic and consists of ferromagnetic columns along the $z$-axis arranged antiferromagnetically in the $xy$-plane. We refer to this GS as a columnar antiferromagnet (CAFM). \\
(III) For $0.127 < \Gamma < 0.164$, the nn exchange favors ferromagnetic alignment and the dipolar interaction continues to dominate. The GS structure changes to a layered antiferromagnet (LAFM) with a {\it two-component} order parameter. \\
(IV) For $\Gamma > 0.164$, the exchange interactions are dominant and the GS is a ferromagnet (FM).

It is interesting to note the systematic transition from the  AFM phase to the FM phase via the onset of ferromagnetic alignment -- first along a single-site one-dimensional column (CAFM), then along a two-dimensional layer (LAFM) and finally . The lower panel depicts the corresponding ground state (GS) configurations for each phase. The symmetry of the Hamiltonian under $\sigma_i \rightarrow -\sigma_i$ spins results in degenerate GS. The degeneracy is {\it two} for Phases I, II, IV, and {\it four} for Phase III. The equilibrium states for a given value of $\Gamma$ at non-zero $T$ will be combinations of the corresponding degenerate GS, separated by frustrated spins due to conflicting interactions. The free energy landscape is complex with several local minima, especially in the dipolar-dominated regimes.

A large number of experimental realizations of the DIM indeed exhibit a non-trivial organization of dipoles in the (so-called) equilibrium phase. For example, magnetic measurements of rare earth hydroxides reveal FM ordering for  $\text{Tb(OH)}_3$, $\text{Dy(OH)}_3$ and $\text{Ho(OH)}_3$, and a complex AFM ordering for  $\text{Nd(OH)}_3$ and $\text{Gd(OH)}_3$ \cite{wolf}. Amongst rare-earth fluorides, LiHoF$_4$ and LiTbF$_4$ exhibit FM ordering, while LiErF$_4$ shows AFM ordering \cite{battison, biltmo09, holmes73, beauvillain}. Similar observations have been made in the context of rare earth ethyl sulphates \cite{Cooke59}, rare earth perovskites \cite{wu2017, wu2019}, and rare earth garnets \cite{cai2019}. The above examples have captured much attention in modern solid state physics and materials science because they exhibit interesting magnetic, multiferroic, and optical effects. Some of these systems, due to their low critical temperatures, are also realizations of quantum Hamiltonians and exhibit phase transitions driven by quantum fluctuations \cite{bitko1996, schechter2005}. The presence of a glassy phase due to frustrated moments arising from conflicting FM and AFM interactions has also been seriously contemplated \cite{piatek2013, andersen2014, Alonso2017}. Recent inclusions in the family of dipolar solids are self-assembled super-lattices of mono-disperse magnetic nanoparticles such as ferric oxide (Fe$_2$O$_3$) or magnetite (Fe$_3$O$_4$) \cite{bae2007, singh2014}. They exhibit a rich phase diagram revealing a variety of stable structures such as hexagonal close-packed, face centered cubic, spherical, cylindrical, etc \cite{talapin2007, boles2016}. Complex and almost perfect geometric ordering of the nano-dipoles has also been observed in microscopy snapshots \cite{Mehdizadeh2015}. Both these observations have been interpreted in the context of dipolar interactions between individual particles.

Laboratory experiments generally require application of external fields that drive the system out-of-equilibrium. The system re-equilibrates, and the approach to equilibrium critically depends on the free energy landscape. An important non-equilibrium study in this context is the kinetics of domain growth or phase ordering, initiated by a sudden quench of the system from the disordered phase to the ordered phase \cite{puribook}. The subsequent domain growth or {\it coarsening}, characterized by a growing length scale $\ell(t)$, is monitored with time. The growth law reveals details of the free energy landscape and relaxation time-scales  \cite{lai1988}. For example, pure and isotropic systems with non-conserved kinetics such as the Ising model ($D=0$) obey the Lifshitz-Allen-Cahn (LAC) law: $\ell(t) \sim t^{1/2}$ \cite{lifshitz1962, allen1979}. It is characteristic of systems with no energy barriers to coarsening and a unique relaxation time-scale. Systems with complicated free energy landscape  due to disorder and competing interactions have a plethora of relaxation time-scales \cite{binder1986,fischer1991,janke2008,ros2019}. The interfaces are usually rough fractals, and the barriers to coarsening grow as a power law of the domain size \cite{villain1984, gaurav2011, gaurav2014}. Domain growth in these systems exhibits logarithmic behavior in the asymptotic limit \cite{corberi2012}.

In spite of its wide-ranging presence, there are only a few studies of the DIM in $d=3$, and non-equilibrium phenomena are even less addressed \cite{bupathy17,Bupathy18}. This is primarily because handling long-ranged interactions is notoriously difficult, both analytically and computationally. Thus challenged, we were motivated to develop theoretical techniques to study the DIM. In a benchmarking study, we investigated coarsening via large-scale Monte Carlo (MC) simulations on cubic lattices $(L=128)$ using Ewald summation procedures to accurately handle the long-range dipolar interactions \cite{bupathy17}. We investigated their effect on growth laws in the FM phase for a limited range of $\Gamma$ values $(\gtrsim 0.16)$. Encouraged by the unusual observation of anisotropic growth laws in this simplest phase, we now undertake the task of exploring the entire phase diagram. Our investigations are guided by the following questions: Are the growth laws distinct in the four phases or are they universal? What information can we obtain about the energy landscapes of the four phases? Is the system characterized by a universal scaling behavior or is it phase specific? Does the complexity and frustration introduced by dipolar interactions yield rough fractal interfaces?

There are two important results in our paper. Our first finding is that, although the equilibrium states in the four phases have distinct symmetry and the interactions are anisotropic, the system obeys the LAC domain growth law $\ell(t)\sim t^{1/2}$ across phases and directions. The growth exponent $1/2$ is {\it universal} for the DIM. The second important result is that the two-point equal-time correlation function that quantifies domain growth exhibits generalized scaling in all phases: $C\left(x,y,z,t\right)= g(\rho/\ell_{\rho}, z/\ell_z)$, where $\rho$ is the radial coordinate in the $xy$-plane. The coarsening system is thus characterized by unique but distinct length-scales along $xy$ and $z$ directions. The scaling function in both directions is {\it universal}, and can be approximated by the well-known Ohta-Jasnow-Kawasaki (OJK) function \cite{ojk,puribook}. The rest of the paper is organized as follows. In Section 2,  we present detailed numerical results on coarsening in phases 1-4  of the DIM. Section 3 provides a summary and discussion.

\section{Coarsening Studies}
\label{coarse}

We now proceed to present detailed results from our studies of coarsening. These studies use local single-spin flip moves which are computationally expensive in contrast to the cluster algorithms often used for equilibrium studies. Further, in systems with long-range dipolar interactions, spins separated by multiple lattice spacings also contribute to the energy. An explicit evaluation of the latter is only possible for small systems as the number of computations scales as $O(N^2)$. A successful procedure for dealing with long-range interactions is the {\it Ewald summation technique} \cite{frenkelbook}. The basic idea here is to write the potential in two parts using the identity:
\begin{equation}
\dfrac{1}{r^3} = \dfrac{f(r)}{r^3} + \dfrac{1-f(r)}{r^3}.
\end{equation}
Here, $f(r)$ is an appropriate splitting function with the following properties: (i) The first part is negligible beyond a certain cutoff $r_{\rm cut}$ so that the summation up to this cutoff is a good approximation of this contribution to the total energy. (ii) The second part is a slowly varying function for all $r$ so that its Fourier transform can be represented by only a few $k$-vectors with $|k| \leq k_{\rm cut} = 2\pi/r_{\rm cut}$. Therefore, this part can be efficiently evaluated in reciprocal space. There are many choices of $f(r)$ that satisfy the above two conditions, but the usual choice is a complementary error function:
\begin{equation}
\text{erfc}(r) = \dfrac{2}{\sqrt{\pi}} \int_r^{\infty} e^{-t^2}dt.
\end{equation}
In particular, we use $f(r) = \text{erfc}(\sqrt{\eta}~r)$, where the Ewald splitting parameter $\eta$ decides the relative weights of the real and Fourier terms. By a suitable choice of $\eta$, we can optimize the computation time for a specified error bound. An excellent discussion of this procedure is found in Ref.~\cite{holm2004}.

Using cubic systems of up to $128^3$ spins, we performed deep quenches to $T = 0.5 \ T_c (\Gamma)$ in the ordered phase. The quench locations are marked in the phase diagram in Fig.~\ref{fig1}. The initial state for all quenches was chosen to be a random configuration of $\sigma_i = \pm 1$ corresponding to the disordered (paramagnetic) phase. The Ewald summation technique with metallic boundary conditions was used to compute the dipolar term of Eq.~(\ref{eq:ham}) \cite{frenkelbook} . We chose the Ewald splitting parameter $\eta = 0.032$, which yields an error of $\delta = 10^{-3}$ in the evaluation of the dipolar term. We also performed simulations designed to yield an error of $\delta = 10^{-4}$ on smaller lattices. This did not alter the growth laws obtained with $\delta = 10^{-3}$ for larger lattices.

The system evolution was studied using spin-flip Glauber dynamics with the standard Metropolis procedure up to 1024 MC steps (MCS). In each phase, the acceptance rate (fraction of spins flipped in 1 MCS) decreases exponentially with time. In Table~\ref{tab1}, we show typical values of the acceptance rate in the 4 quenches studied in this paper.
\begin{table}[h]
\begin{center}
\caption{Typical values of the acceptance rate at different times in the four quenches studied here.}
\label{tab1}
\begin{tabular}{|l|c|c|c|c|}
\hline
$\Gamma$ & $-10.0$ & 0.0 & 0.14 & $\infty$ \\
\hline
$t=10^0$ & 0.345 & 0.358 & 0.335 & 0.537 \\
\hline
$t=10^1$ & 0.052 & 0.080 & 0.066 & 0.147 \\
\hline
$t=10^2$ & 0.014 & 0.024 & 0.027 & 0.055 \\
\hline
$t=10^3$ & 0.001 & 0.003 & 0.003 & 0.012 \\
\hline
\end{tabular}
\end{center}
\end{table}
All statistical quantities have been averaged over $10$ different initial conditions. This results in error bars for the correlation function and length scale data which are smaller than the symbol sizes we use subsequently for these quantities.

\begin{figure}
\begin{center}
\includegraphics[width=14cm]{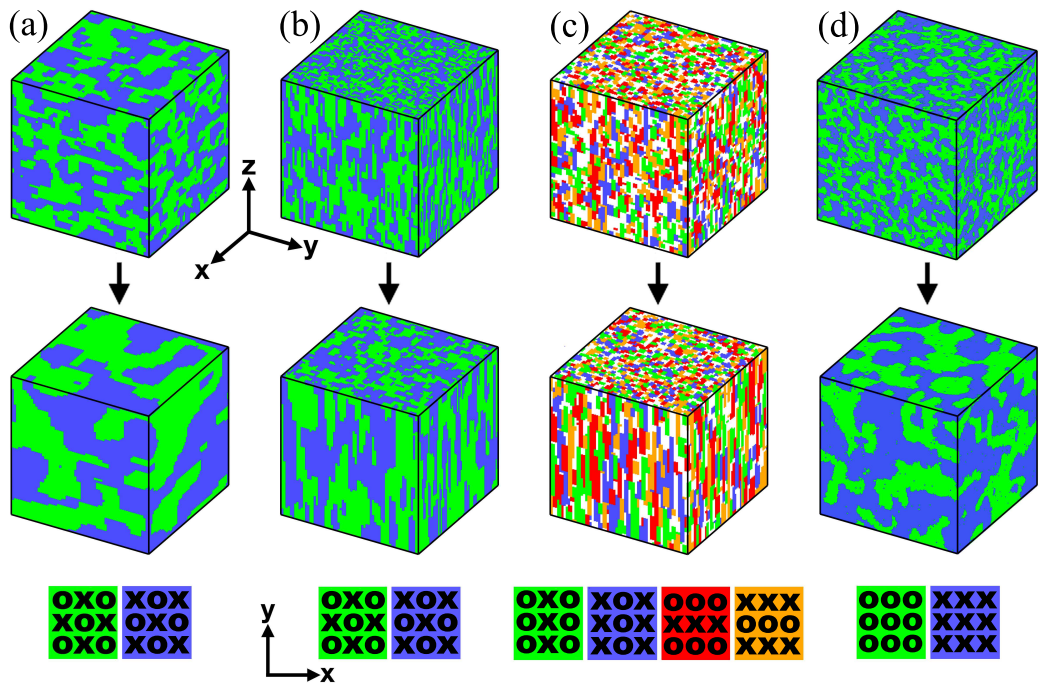}
\end{center}
\caption{Evolution snapshots of the staggered magnetization at $t=8$ MCS (top row) and $t= 64$ MCS (second row) for the four phases: (a) AFM, (b) CAFM, (c) LAFM, and (d) FM. The key shows the color code for the degenerate GS in each phase, as described in the text.}
\label{fig2}
\end{figure}

In Fig.~\ref{fig2}, we present typical snapshots of coarsening morphologies in the four phases. The top row shows the snapshots at $t=8$ MCS, and the second row at $t= 64$ MCS. During the coarsening process, the degenerate equilibrium states compete with one another, and the corresponding domains are separated by interfacial defects. As time evolves, the defects annihilate and the system selects one of the ground states. The different colors in each snapshot represent domains of one of the degenerate ground states of that phase, as shown in the keys below. To identify the domains in phases I-III, a standard prescription is used to define the staggered magnetization: \\
(i) In phase I, the morphology has AFM spin arrangement along $x$, $y$ and $z$ directions. The staggered spin variables are obtained using:
\begin{equation}
\psi_{xyz} = (-1)^{x+y+z} \sigma_{xyz}.
\label{p1_stag}
\end{equation}
The green (blue) domains in Fig.~\ref{fig2}(a) correspond to correlated regions of up (down) spins in the staggered representation. The key below also shows the corresponding GS configurations in terms of the original spin variables $\sigma_{xyz}$. \\
(ii) In phase II, the morphology is AFM along the $x$ and $y$ directions, and FM along the $z$ direction. So the spins are staggered using:
\begin{equation}
\psi_{xyz} = (-1)^{x+y} \sigma_{xyz} .
\label{p2_stag}
\end{equation}
The green (blue) domains in Fig.~\ref{fig2}(b) correspond to correlated regions of up (down) spins in the staggered representation of the GS. \\
(iii) In phase III, the morphology comprises of FM $xz$ (or $yz$) planes with an AFM arrangement in the $y$ ($x$) direction. In such morphologies, the system is characterized by a two-component order parameter  $\vec{\psi}_{xyz}\equiv\left(\psi^1_{xyz},\psi^2_{xyz}\right)$ \cite{krinsky, sadiq} with
\begin{eqnarray}
\psi^1_{xyz} &=& (-1)^{x} \frac{1}{4}\left[\sigma_{xyz} + \sigma_{x(y+1)z} - \left(\sigma_{(x+1)yz} + \sigma_{(x+1)(y+1)z}\right)\right] , \\
\psi^2_{xyz} &=& (-1)^{y} \frac{1}{4}\left[\sigma_{xyz} + \sigma_{(x+1)yz} - \left(\sigma_{x(y+1)z} + \sigma_{(x+1)(y+1)z}\right)\right] .
\label{p3_stag}
\end{eqnarray}
In the morphologies in Fig.~\ref{fig2}(c), a site is assigned: green for $\vec{\psi}_{xyz} = (1,0)$; blue for $\vec{\psi}_{xyz} = (-1,0)$; red for $\vec{\psi}_{xyz} = (0,1)$; orange for $\vec{\psi}_{xyz} = (0,-1)$. Each of the colored domains represents correlated regions corresponding to one of the four GS, as shown in the key below. \\
(iv) No staggering is required for the morphologies in Fig.~\ref{fig2}(d) in the FM phase. For uniformity of notation, we assign $\psi_{xyz} = \sigma_{xyz}$. \\
Clearly, the domains in each phase grow with time, although the growth is faster in the $z$ direction as compared in the $xy$ plane. However, the growth in the $xy$ plane for Fig.~\ref{fig2}(c) is arrested as the system gets stuck in metastable states. The problem of metastability is widespread in phase III due to the presence of four competing GS.  

The length scales associated with the evolving domains can be evaluated from the two-point equal-time correlation function:
\begin{equation}
C(\vec{r},t) = \langle \psi (\vec{r}_1,t) \psi(\vec{r}_2,t) \rangle - \langle \psi(\vec{r}_1,t)\rangle \ \langle\psi(\vec{r}_2,t)\rangle .
\end{equation}
Here, $\psi$ is the appropriate order parameter, $\vec{r} = \vec{r}_1 - \vec{r}_2$, and $\langle \cdot\cdot\cdot \rangle$ represents an average over independent runs \cite{puribook}. In the case of isotropic domain growth, the {\it dynamical scaling ansatz} assumes the presence of a single length scale $\ell (t)$. This is demonstrated post facto by the scaling collapse of the correlation function: $C(\vec{r},t) = g(r/\ell)$. The validity of this has been shown in many experiments and simulations. In the anisotropic case considered here, we propose (following Ref.~\cite{pbd92,ppd94}) the simplest anisotropic generalization of this ansatz. This also has to be verified post facto by the dynamical scaling of the correlation function. Thus, we introduce $C(\vec{r},t) \equiv C(\vec{\rho},z;t)$ where $\vec{\rho} = (x,y)$. In the case of unique length scales $\ell_{\rho}$ and $\ell_z$ characterizing domain growth in the $xy$ and $z$-directions, the correlation function should exhibit {\it generalized dynamical scaling}: $C(\vec{\rho},z,t) = g(\rho/\ell_{\rho},z/\ell_z)$.

In the isotropic case, an approximate analytical form of the correlation function for a system described by a scalar nonconserved order parameter has been obtained by Ohta et al. (OJK) by studying the interfacial defect dynamics \cite{ojk,puribook}. The OJK function is given by
\begin{equation}
C(r,t)= \frac{2}{\pi} \sin^{-1}\gamma, \quad \gamma = \exp\left(-r^2/\ell^2\right) .
\end{equation}

\begin{figure}
\begin{center}
\includegraphics[width=14cm]{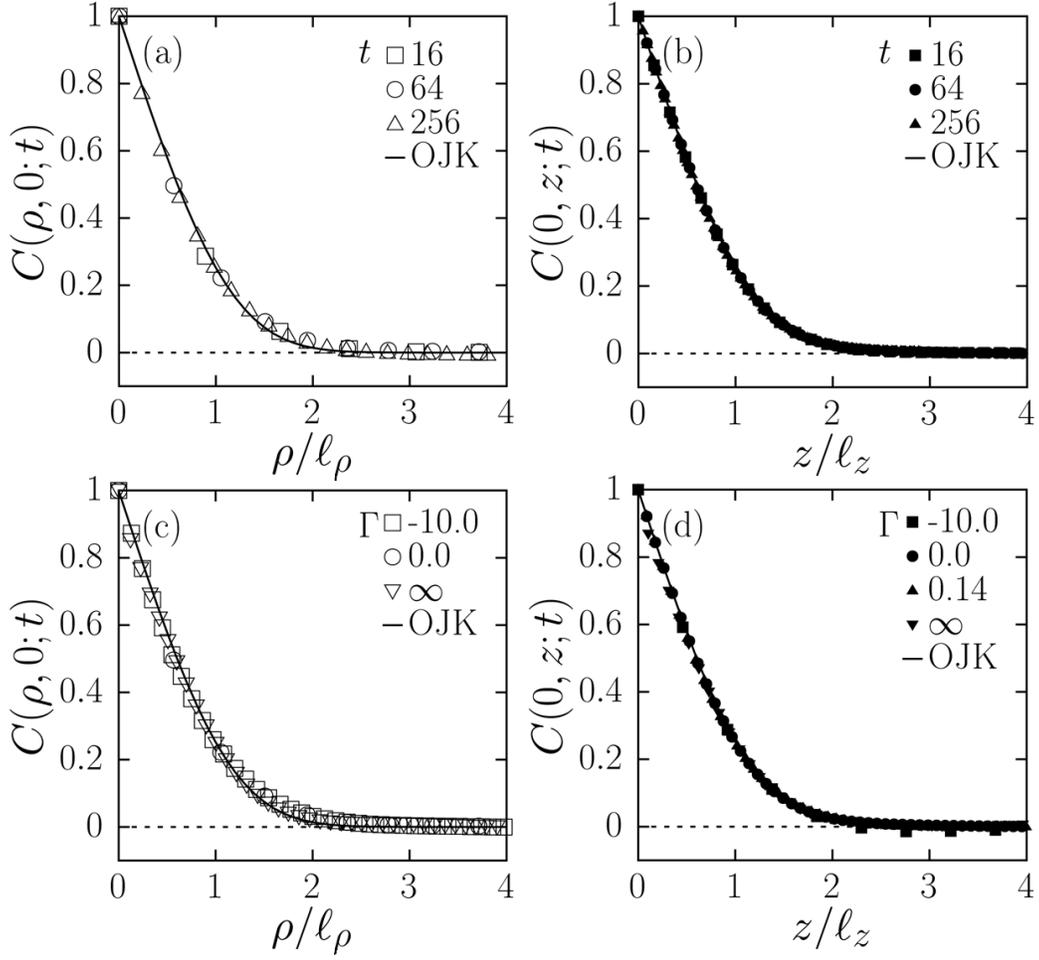}
\end{center}
\caption{Scaled correlation functions at $t=16,64,128$ for $\Gamma = 0$ in phase II: (a) $C(\rho,0,t)$ vs. $\rho/\ell_{\rho}$, and (b) $C(0,z,t)$ vs. $z/\ell_z$. Scaled correlation functions at $t=64$ for specified $\Gamma$-values: (c) $C(\rho,0,t)$ vs. $\rho/\ell_{\rho}$, and (d) $C(0,z,t)$ vs. $z/\ell_z$. The solid line in (a)-(d) represents the OJK function.}
\label{fig3}
\end{figure}

In Fig.~\ref{fig3}, we show the scaled correlation functions for phase II with $\Gamma=0$: $C(\vec{\rho},0,t)$ vs. $\rho/\ell_{\rho}$ in Fig.~\ref{fig3}(a), and $C(0,z,t)$ vs. $z/\ell_z$ in Fig.~\ref{fig3}(b). The length scales are self-consistently defined from the data as the distance over which the relevant correlation function decays to 0.2 times its maximum value. We emphasize that no fitting parameters have been used to observe the data collapse. Both data sets exhibit dynamical scaling, indicating that the system is characterized by unique but distinct length-scales along $xy$ and $z$-directions.  More striking is their agreement with the OJK function (solid line), which shows that the defect dynamics is {\it robust} across directions in spite of the inherent anisotropy of the dipolar interactions. Fig.~\ref{fig3}(c) shows $C(\vec{\rho},0,t)$ vs. $\rho/\ell_{\rho}$ for $t = 64$ in phases I-IV for the specified values of $\Gamma$. Fig.~\ref{fig3}(d) shows the corresponding data for $C(0,z,t)$ vs. $z/\ell_z$ for $t = 64$. The data collapse is excellent in both Figs.~\ref{fig3}(c)-(d), and is well-approximated by the OJK form. These observations reveal that the coarsening morphologies are {\it robust} across diverse phases of the DIM. 

Finally, we discuss the growth laws in the four phases of the DIM. In this context, it is useful to reiterate observations from earlier studies \cite{bray3} of domain growth in the Ising model with {\it isotropic} long-range interactions:
\begin{equation}
\label{LIM}
J(r_{ij}) \sim r_{ij}^{-(d+\mu)}. 
\end{equation}
Using energy scaling arguments, Bray-Rutenberg \cite{bray3}  predicted the following domain growth laws:
\begin{equation}
\label{GL_LIM}
\ell(t) \sim  \left\{\begin{array}{ll}
t^{1/1+\mu}, &\mu <1,\\
(t\ln t)^{1/2}, & \mu=1,\\
t^{1/2},            & \mu>1. \\
\end{array}
\right.
\end{equation}
A recent work by Christiansen et al. \cite{suman2019} has studied coarsening in the $d=2$ Ising model with additional long-range interactions as in Eq.~(\ref{LIM}). Using efficient numerical schemes, the authors confirmed the predictions in Eq.~(\ref{GL_LIM}) for many values of $\mu$.

The angular dependence of the dipolar interactions yields anisotropic morphologies due to ferromagnetic interactions along the $z$-direction and antiferromagnetic interactions in the $xy$-plane. Roughly speaking, the system behaves as a long-range Ising model along the $z$-axis with $d=1,\mu=2$. Then, the expected growth law from Eq.~(\ref{GL_LIM}) is $\ell_{z}(t)\sim t^{1/2}$. On the other hand, in the $xy$-plane, we have $d=2,\mu=1$ in Eq.~(\ref{LIM}). Again, we expect the growth to be predominantly diffusive with logarithmic corrections.

\begin{figure}
\begin{center}
\includegraphics[width=14cm]{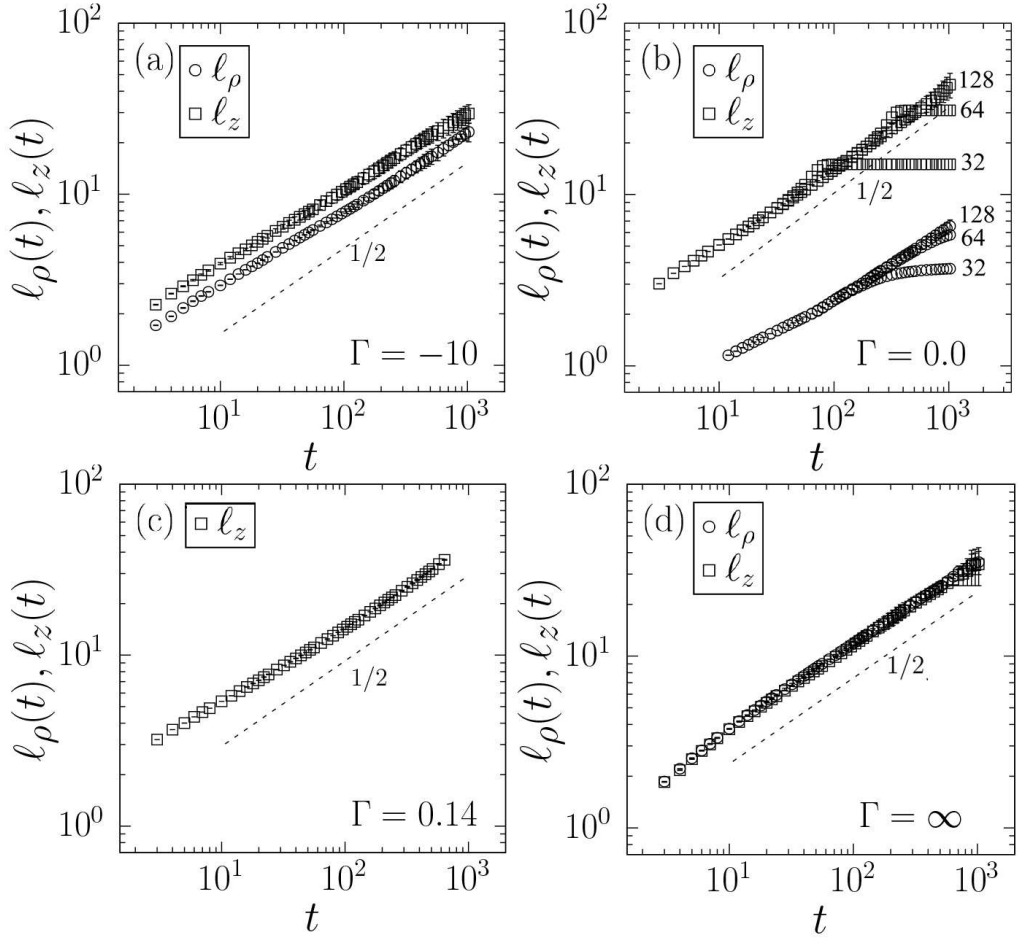}
\end{center}
\caption{Characteristic length scales, $\ell_\rho(t)$ vs. $t$ and $\ell_z(t)$ vs. $t$ on a log-log scale. The panels correspond to (a) $\Gamma=-10$; (b) $\Gamma=0$; (c) $\Gamma=0.14$; (d) $\Gamma=\infty$. In (b), we also include data for systems of size $32^3$ and $64^3$ to show the finite-size effects. In (c), we have not shown $\ell_\rho$ vs. $t$ as there is almost no growth in that direction. In (d), the data sets for $\ell_\rho$ and $\ell_z$ are numerically indistinguishable as the growth is isotropic. The dashed line in each panel denotes the LAC law: $\ell(t) \sim t^{1/2}$.}
\label{fig4}
\end{figure}

In Fig.~\ref{fig4}, we present the growth laws $\ell_{\rho}(t)$ vs. $t$ and $\ell_{z}(t)$ vs. $t$ on a log-log scale for phases I-IV. We have not shown $\ell_{\rho}$ vs. $t$ for the LAFM phase as the growth in the $xy$ plane is almost frozen in that phase \cite{kretschmer}. Further, as the growth in the $z$-direction is very rapid, we see finite-size effects in the late-time data for the FM phase. The error bars for each data point are smaller than the symbol sizes used except in the regime where finite-size effects are seen. The dashed lines with the expected slope 1/2 have been plotted for reference. It is striking that the growth in both directions, in spite of the diversity of phases and the inherent anisotropy of the dipolar interactions, obeys the universal LAC law $\ell(t)\sim t^{1/2}$! Their presence is thus confined to the prefactors alone. There are no length-scale-dependent barriers to coarsening in the DIM, and the non-equilibrium evolution is characterized by a unique relaxation time-scale. (In our earlier study, the domain growth in the FM phase is studied for smaller values of $\Gamma\gtrsim 0.16$ that are close to the phase boundary separating the LAFM and FM phase \cite{bupathy17}. Here, longer timescales are required to observe the $t^{1/2}$ law and the anisotropy is again contained in the pre-factors.)

Before concluding, it is relevant to ask how the finite size of the system would affect the growth laws. In contrast to the equilibrium case, various physical quantities do not change systematically with the system size. Rather, the domains grow in a power-law manner until the domain scale becomes some significant fraction of the lateral system size. After that, finite-size effects are seen via a crossover to a regime with flattening and saturation of the domain growth law. Typically, the data sets for different system sizes are numerically coincident until they encounter finite-size effects. This is shown in Fig.~\ref{fig4}(b) for the case with $\Gamma = 0$.

\section{Summary and Discussion}

We end this paper with a summary of our results, their implications, and future directions. We have performed large-scale Monte Carlo (MC) simulations to study coarsening dynamics in the dipolar Ising model (DIM) which encompasses short-range exchange interactions as well as long-range dipolar interactions. This model is characterized by four distinct phases: (i) {\it antiferromagnet} (AFM), (ii) {\it columnar antiferromagnet} (CAFM), (iii) {\it layered antiferromagnet} (LAFM), (iv) {\it ferromagnet} (FM). The dipolar interactions lead to diverse ground state (GS) configurations with strong anisotropy. Yet the non-equilibrium dynamics is characterized by universality. Our main observations are (i) the spatial correlation function exhibits universal scaling; (ii) The domain growth law obeys the universal Lifshitz-Allen-Cahn law $\ell(t) \sim t^{1/2}$.

Thermal quenches are a starting point for many non-equilibrium studies in the laboratory. In these experiments, the system often accesses long-lived metastable states that are encountered in our simulations. So our observations could provide a fresh outlook to interpret relaxation phenomena in dipolar solids \cite{chamberlain2005, biltmo09, wu2019}. Our work also suggests novel experimental investigations in the more contemporary self-assembled super-lattices \cite{ku2011}. In the latter system, the constituent magnetic nanoparticles are usually functionalized with an insulating surfactant layer to prevent aggregation. The surfactant thickness can be adjusted to manipulate dipole-dipole interactions to tailor spin morphologies dictated by applications. For example, a large class of self-assembled lattices finds applications for spintronic devices that require an AFM arrangement of the nano-dipoles for efficient operation \cite{black2000, kechrakos2005, dugay2011, usmani2018, arun2019, jardin2011, manish2019}. An improved understanding of the interplay of short-range and long-range interactions would yield better strategies to achieve such challenges. 

To the best of our knowledge, this is the first study of coarsening in ($d=3$) dipolar solids. The developed methodologies provide a basis for relaxation studies in systems with anisotropic and long-range interactions in general. We have identified unexpected dynamical universalities in the DIM, which is representative of a large class of microscopic and mesoscopic systems. However, there are many puzzles that still need to be understood. For example the basis for universality, the role of lattice geometry and other forms of long-range interactions, and the possibility of the glassy state in the LAFM phase are open questions. Another non-equilibrium phenomenon of great relevance in experimental systems is that of {\it aging} \cite{binder1986,fischer1991}. This property probes the history-dependent evolution of correlation and response functions when the system is driven out-of-equilibrium by, e.g., a thermal quench or the application of a magnetic field. There are a few such studies for the Ising model with long-range interactions in $d=2$, and these have suggested novel violations, phases and exponents \cite{cannas1999,cannas2008,suman2020}. It will be interesting to study aging phenomena in the anisotropic DIM. The consequence of special directions on the relationship between waiting times and relaxation times is an open question. We hope that our study will motivate investigations to seek these answers, which are important for fundamental understanding as well as experimental interpretations.

\subsection*{Acknowledgements}
SK acknowledges the HPC facility at the Jawaharlal Nehru University, New Delhi, and useful discussions with Arunkumar Bupathy. VB acknowledges DST, India for a MATRICS research grant.

\bibliography{dgadim.bib}

\end{document}